\documentclass[fleqn,usenatbib]{mnras}

\usepackage{newtxtext,newtxmath}

\usepackage[T1]{fontenc}

\DeclareRobustCommand{\VAN}[3]{#2}
\let\VANthebibliography\thebibliography
\def\thebibliography{\DeclareRobustCommand{\VAN}[3]{##3}\VANthebibliography}

\usepackage{graphicx}	
\usepackage{amsmath}	
\usepackage{tabularx}

\title{Detection of Five New RRATs at 111 MHz}

\author[S. A. Tyul'bashev et al.]{
S. A. Tyul'bashev $^{1}$\thanks{E-mail: serg@prao.ru}
V.S. Tyul’bashev,$^{1}$
V. M. Malofeev,$^{1}$
S. V. Logvinenko,$^{1}$
V. V. Oreshko,$^{1}$\newauthor 
R. D. Dagkesamanskii,$^{1}$
I. V. Chashei,$^{1}$
V. I. Shishov,$^{1}$
N. N. Bursov$^{2}$
\\
$^{1}$ Pushchino Radio Astronomy Observatory, Astro Space Center, Lebedev Physical Institute, Russian Academy of Sciences, Pushchino, 142290 Russia\\  
$^{2}$ Special Astrophysical Observatory, Russian Academy of Sciences,
Nizhnii Arkhyz, Karachai-Cherkessian Republic, 357147 Russia
\\
}

\date{July 31, 2017}

\pubyear{September 5, 2017}

\begin{document}
\label{firstpage}
\pagerange{\pageref{firstpage}--\pageref{lastpage}}
\maketitle

\begin{abstract}
Results of 111-MHz monitoring observations carried out on the Big Scanning Antenna of the Pushchino Radio Astronomy Observatory during September 1–28, 2015 are presented.  Fifty-four pulsating sources were detected at declinations $-9^o<\delta<42^o$. Forty-seven of these are known pulsars, five are new sources, and two are previously discovered transients. Estimates of the peak flux densities and dispersion measures are presented  or all these sources. 
\end{abstract}




\section{Introduction}
The detection of 11 objects with irregular, short flares at 1.4 GHz was reported in  2006 (\citeauthor{McLaughlin2006}, \citeyear{McLaughlin2006}). These rapid, very short pulses with durations from 2–30 ms were found in  archival records of a Parkes pulsar survey. They displayed a clear shift in arrival time as a function of the observing frequency, which outwardly resembled individual  pulses from strong pulsars. When a sufficient number of repeating pulses are detected from a single direction, it was possible to carry out timing to measure the period of the pulsar. Since the high density of the pulses at first made them seem different from ordinary pulsar emission, such objects were given the name Rotating RAdio Transients, or RRATs. Thus, RRATs represent a special case of pulsar activity. The first RRAT discovered at the Pushchino Radio Astronomy Observatory (PRAO) was found at 111 MHz in observations taken in 2004 and 2007 (\citeauthor{Shitov2009}, \citeyear{Shitov2009}), i.e., at about the same time.

In addition to RRATs, flares from extragalactic sources were also detected at that same time. Such flares can be indistinguishable from the pulses of RRATs in terms of both duration and their dispersion measure. An example of a single stong pulse with a dispersion measure of 375 pc/cm$^3$ detected in the Parkes archival records was  
presented in the 2007 paper \citeauthor{Lorimer2007} (\citeyear{Lorimer2007}). Because the dispersion measure exceeds the expected value of 45 pc/cm$^3$, this flare was interpreted as being extragalactic, with the  expected distance to the source of the flare being of order a billion light years. Such flares are called Fast Radio Bursts (FRBs). RRATs and FRBs are special cases of fast transients. In contrast to RRATs, extragalactic FRBs are mainly observed only once, apart from the single known case of a repeating FRB (\citeauthor{Spitler2016}, \citeyear{Spitler2016}).

Recent especially organized searches for fast transients, including those contained in archival records, have led to the detection of about 100 such objects, most of which have repeating flares (see the ATNF pulsar catalog ${ http://www.atnf.csiro.au/people/pulsar/psrcat/}$ and the RRAT catalog (RRATalog)${ http://astro.phys.wvu.edu/rratalog/}$). Most of the objects in these catalogs are probably RRATs. It is interesting that, according to \citeauthor{Keane2010} (\citeyear{Keane2010}), the number of detectable pulsars without a clearly expressed period could appreciably exceed the number of classical radio pulsars. Since the ATNF catalog already contains more than 2600 pulsars, we expect that a specialized long-term search for RRATs could yield a large number of new detections.

In an ordinary pulsar search, it is possible to obtain a long recording of the signal, making it possible to accumulate a large number of pulses. Since the final signal-to-noise ratio (SNR) is proportional to the square root of the number of pulses, this makes it possible to realize high sensitivity in a pulsar search
using relatively modest radio telescopes. A high fluctuational (instantaneous) sensitivity is needed for searches for fast transients, making it necessary to
use large telescopes. The remaining required conditions are the same as those for pulsar searches: multifrequency recording of the observations and the use of a time constant or readout time comparable to the duration of the flares. All these conditions are satisfied in our monitoring survey carried out at 111 MHz at the PRAO. The reduction of five days of observations from this survey yielded the registration of flares from fast transients in two directions. In one of these directions, flares were detected over several days (\citeauthor{Tyulbashev2017}, \citeyear{Tyulbashev2017}). Here, we present and discuss the results of a set of observations with a duration of four weeks.

\section{OBSERVATIONS AND REDUCTION}

Regular monitoring observations on the PRAO Big Scanning Antenna (BSA) began in 2013 at a central frequency of 110.25 MHz in a regime with six frequency channels with a total bandwidth of 2.5 MHz and a readout time of 100 ms. Starting in August 2014, these have been carried out in parallel with observations with high frequency and time resolution: 32 frequency channels with a time constant of 12.5 ms. The BSA is a multi-beam antenna, and can operate as several different radio telescopes. In particular, for monitoring, we use a 128-beam radio telescope whose beams cover declinations from $-9^o$ to $+55^o$. At present, the digital radiometers are connected to 96 beams of this radio telescope, corresponding to declinations from $-9^o$ to $+42^o$. The beams overlap at the 0.4 level (see \citeauthor{Shishov2016} (\citeyear{Shishov2016}), \citeauthor{Tyulbashev2016} (\citeyear{Tyulbashev2016}) and the site http://bsa-analytics.prao.ru for more detail), with the size of a single beam being $\simeq 0.5^o \times 1^o$.

The effective area of the radio telescope in the direction toward the zenith is 45 000–50 000 m$^2$. Taking into account the receiver bandwidth, time constant, and antenna temperature, which we took to be 1000 K, we can estimate the fluctuational sensitivity of the BSA toward the zenith in its 32-channel frequency regime to be $S_{fl} = 0.3$~Jy. This sensitivity could be a factor of a few times worse when the position of a source on the sky is unfavorable relative to the fixed beams of the antenna, for example, when the source falls between two beams.

To search for fast transients in the monitoring data, we added the frequency channels applying DM = 3–100~pc/cm$^3$. A large number of false sources arise for DM < 3~pc/cm$^3$, and a specialized search method must be used to detect transients with such low dispersion measures. The detection of transients at large dispersion measures is also problematic. Our pulsar search at 111 MHz shows that the total number of detected known and new pulsars falls rapidly with increasing dispersion measure. We were not able to detect a single new pulsar at dispersion measures greater than 80~pc/cm$^3$ [8, 9] (\citeauthor{Tyulbashev2016} (\citeyear{Tyulbashev2016}), \citeauthor{Tyulbashev2017a} (\citeyear{Tyulbashev2017a})). In our opinion, this is due to the fact that the broadening of a pulse in a frequency channel is already comparable to the readout interval with dispersion measures greater than 25~pc/cm$^3$. In addition, scattering is large at low frequencies, decreasing the peak flux density. Therefore, the search for transients was initially restricted to a maximum dispersion measure of 100~pc/cm$^3$.

In a test search for transients in a recording accumulated taking into account the assumed dispersion measure, the threshold SNR was taken to be 7. This corresponds to a detection threshold of 2.1 Jy at the pulse maximum at 111 MHz. The level of  interference mainly corresponds to SNRs less than 7, and we accordingly specified a threshold SNR of 7 in our search. A reliable visual identification of transients
can be carried out in the remaining recordings.

\begin{table*}
\centering
\caption{Comparison of threshold levels for detecting transients on various telescopes.}
\label{tab:tab1}
\begin{tabular}{cccc}
\hline
Telescope & $\nu_{center}$ (MHz) & $S_{peak}$ (Jy) & $S_{peak}$ (Jy) \\
\hline
Parkes 64-m      & 1400 & 0.1 (\citeauthor{McLaughlin2006} (\citeyear{McLaughlin2006}))   & 4\\
Green Bank 100-m & 350  & 0.1 (\citeauthor{Karako-Argaman2015} (\citeyear{Karako-Argaman2015}))  & 0.55\\
Arecibo 300-m    & 327  & 0.1 (\citeauthor{Deneva2016} (\citeyear{Deneva2016}))  & 0.5\\
PRAO BSA         & 111  & -         & 2.1\\
\hline
\end{tabular}
\label{tab:tab1}
\end{table*}

Let us compare the minimum peak flux densities of transients found on various large telescopes and the expected threshold detection level for radio transients for the BSA in the zenith direction. The columns of Table~\ref{tab:tab1} present (1) the telescope, (2) the central frequency at which transient searches have been carried out, (3) the estimated peak flux density and the reference from which it was taken, and (4) a recalculation of the values from column (3) to 111 MHz assuming a spectral index of $\alpha=1.5 (S\sim \nu^{-1.5})$, close to the mean spectral index in the range 102–400 MHz (\citeauthor{Malofeev2000} \citeyear{Malofeev2000}).

The real 111 MHz threshold levels will be worse than the estimates in Table 1 by a factor of two to three, on average. This is due to the fact that most observations are not carried out at the zenith, it is unlikely that a source will fall precisely into one of the beams, and the time constant and width of the frequency channels are not optimal for transient searches. Thus, the sensitivity of the BSA is lower than the sensitivities of two of the three other largest radio telescopes used in searches for individual pulses. Taking into account the peak flux densities of transients published in these other studies, this difference could comprise nearly an order of magnitude in the case of an unfavorable position of the source relative to one of the BSA beams.

In reality, the situation is not quite so bad. With regard to pulsars, it is known that the observed flux densities of individual pulsar pulses depend on both internal and external factors. One internal factor is the intrinsic variability of the emission. An external factor that influences the apparent flux density of a pulse is interstellar scintillation. The influence of these factors is described in our search for pulsars using the same monitoring data (\citeauthor{Tyulbashev2017a}, \citeyear{Tyulbashev2017a}), and we do not consider them further here.

There are also additional factors leading to possible changes in the received pulse flux density for fast transients. First, roughly $75\%$ of the observed area
falls in the zone that is optimal for interplanetary scintillations, with elongations 20$^o$-40$^o$. Since transients are essentially point sources, they should scintillate on the interplanetary plasma, and we expect that the relative variations of the flux densities of two individual pulses that “favorably” and “unfavorably” hit moving inhomogeneities in the interplanetary plasma could strengthen or weaken an observed pulse by a factor of 1.5–2. Second, the overall energy distribution of pulsar pulses shows that, statistically, a pulse that is appreciably stronger than the mean pulse can always be encountered in sufficiently long observations. If giant pulses are observed for a given pulsar, the tail of the distribution becomes power-law, rather than log-normal (see, e.g., \citeauthor{Kazantsev2017} (\citeyear{Kazantsev2017})), and the probability of detecting the pulsar through its individual pulses increases.

Since RRATs are pulsars, we expect that there will always be some probability of detecting a strong pulse from a RRAT in long observations of a single region of sky; in studies aiming to detect RRATs, the reason for the enhancement of a RRAT’s pulse is not important to us. The formal sensitivity of the BSA may be lower than those of other radio telescopes used in RRAT searches, but the RRAT-search sensitivity is effectively higher than those of other surveys due to the long duration of the BSA monitoring. Note that, in monitoring with the BSA, an area of sky roughly equal to 15 000 deg$^2$ is observed daily.

The processing of the monitoring data was performed using a modified version of the BSA-Analytics code used earlier in pulsar searches (\citeauthor{Tyulbashev2016}, \citeyear{Tyulbashev2016}). This code is written in Qt/C++ and is distributed under the license GPLv3. The input files and documentation
are accessible at https://github.com/vtyulb/BSAAnalytics.

The processing was carried out on two cores of a personal computer with a Pentium CPU G680 processor with a frequency of 3 GHz. The processing of one month of observations occupied seven days. A preliminary transient search was run in a singlecore regime with one processor, which made it possible to appreciably reduce the total calculation time. Approximately 10$^7$ sources were distinguished during the operation of this program. After filtering out interference and identifying transients using various criteria (see the following Section), the total number of fast-transient candidates was reduced by a factor of 100. Thus, the output of the program yielded a catalog containing 100 000 sources detected during the month of observations. In our hand reduction of the catalog, we also used a number of additional filters (rejection of known pulsars, of sources detected in more than three beams, and of sources detected at only one dispersion measure, as well as others), which helped substantially reduce the number of candidate fast transients. All remaining sources were verified visually.

Let us now list the main indications of detection of a RRAT: (1) presence of a dispersive pulse in no more than two neighboring spatial beams; (2) presence of a repeating signal at the same sidereal time in all beams; (3) coincidence of the observed dispersion measures from event to event. If there was no repeated detection, the source remained a RRAT candidate.

\section{RESULTS}

In a visual inspection of the results of a blind search for transients, the coordinates and dispersion measures of detected pulses usually identify them with known pulsars. As a rule, many pulses with the same dispersion measure are detected for these pulsars. The pulsar periods are not explicitly visible, since most of the pulses are not directly detected with our adopted SNR criterion. The identification of the pulse coordinates and dispersion measures was carried out using the ATNF catalog. In all, we detected 46 known pulsars. We constructed dynamical spectra and obtained profiles for strong pulses of each pulsar.

The dynamical spectra were constructed using independent channel normalization. In each frequency channel of a dynamical spectrum, the highest flux density in ADC units was assigned the color black, and the lowest flux density white. A 256-level gray scale was then introduced between the two. The $BSA-Analytics$ code is able to apply a running mean filter over the frequency channels, making it possible to enhance the contrast for pulses with widths larger than the readout interval; it is also possible to add frequency channels, making it possible to increase the SNR for sources with relatively modest dispersion measures.

\begin{figure*}
\begin{center}
	\includegraphics[width=0.9\textwidth]{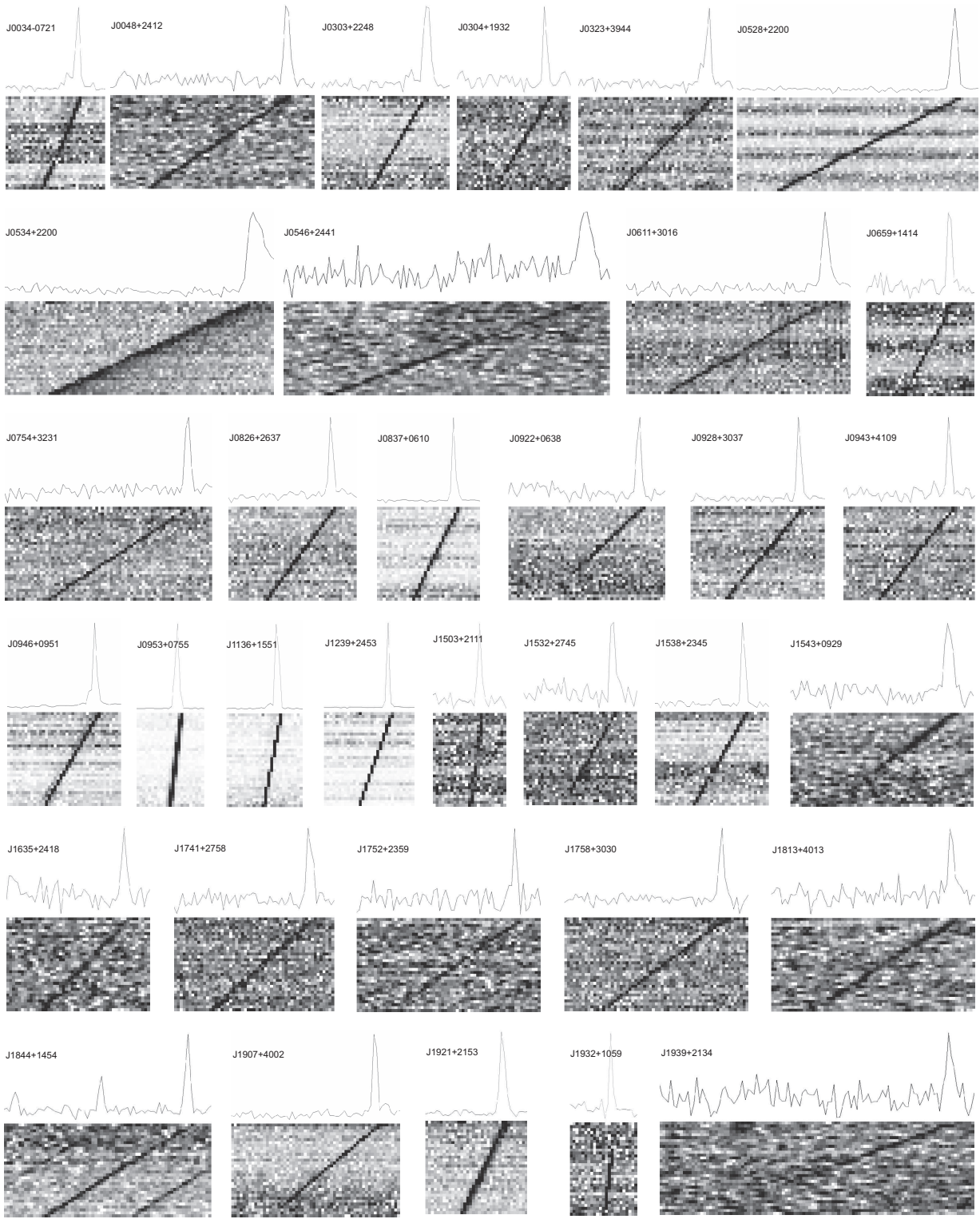}
    \caption{Profiles of individual pulses and dynamical spectra of known pulsars detected in the blind search.}
    \label{fig:fig1atransients}
\end{center}
\end{figure*}

Fig.~\ref{fig:fig1atransients} presents the best recordings of individual pulses for all the previously known pulsars. In the dynamical spectra, the uppermost frequency is 111.046 MHz and the lowermost 109.04 MHz. The horizontal axis covers the duration of the flares with a small margin. No signal is visible in individual frequency channels in some of the dynamical spectra, especially those for declinations below $\delta < +5^o$. This is related to the fact that the BSA is a diffraction array, and different frequencies correspond to different directions on the sky when observing at low declinations. Horizontal stripes are visible in some of the spectra; these could be related to the absence of absolute calibration in each frequency channel (changes in the gain between frequency channels can lead to the appearance of such stripes). On the other hand, the appearance of stripes could also be due to the presence of Faraday rotation. We will examine our results for these pulsars further in subsequent studies.

In Table~\ref{tab:tab2} lists all the pulsars for which individual strong pulses were detected with SNRs greater than 7. The columns present (1) the J2000 name of the pulsar, (2)–(3) the catalog values of the period and dispersion measure, (4) the dispersion measure used to sum the pulses over the frequency bands, (5) the flux density at 400 MHz, (6) the estimated flux density at 102 MHz according to \citeauthor{Malofeev2000} (\citeyear{Malofeev2000}), (7) the upper limit to the pulse half-width, (8) the expected peak flux density in the average pulse at 111 MHz ($S_{peak1}$), (9) the observed peak flux density ($S_{peak2}$) of the pulse presented in Fig.~\ref{fig:fig1atransients}, and (10) the peak flux density recalculated from $S_{peak2}$ by taking into account corrections for the zenith distance and the position of the pulsar between the BSA beams, $S_{peak3}$.

\begin{table*}
	\centering
	\caption{Pulsars detected as fast transients at 111MHz.}
	\label{tab:tab2}
	\begin{tabular}{cccccccccc} 
		\hline
Pulsar name & P (s) & $DM$ pc/cm$^3$ & $DM_{111}$ pc/cm$^3$ & $S_{400}$ mJy & $S_{111}$ mJy & $W_{1/2}$ (ms) & $S_{peak1}$ (mJy) & $S_{peak2}$ (mJy)  & $S_{peak3}$ (mJy)\\
		\hline
J0034-0721	&	0.9429	&	10.9	&	11	&	52	&	560	&	56.7	&	9300	&	8000	&	40000	\\
J0048+3412	&	1.217	&	39.9	&	40	&	2.3	&	88	&	21.7	&	4900	&	4000	&	5000	\\
J0303+2248	&	1.207	&	20	&	19	&	-	&	-	&	50	&	-	&	6000	&	7500	\\
J0304+1932	&	1.3875	&	15.7	&	16	&	27	&	48	&	58.4	&	1100	&	8000	&	16700	\\
J0323+3944	&	3.032	&	26.2	&	26	&	34	&	230	&	42.7	&	16300	&	12000	&	15800	\\
J0528+2200	&	3.7455	&	50.9	&	50	&	57	&	100	&	185.5	&	2000	&	12000	&	15000	\\
J0534+2200	&	0.0333	&	56.8	&	56	&	550	&	10000	&	3	&	-	&	-	&	-	\\
J0546+2441	&	2.8438	&	73.8	&	70	&	2.65	&	-	&	28	&	1800	&	2900	&	3300	\\
J0611+3016	&	1.412	&	45.3	&	45	&	1.4	&	-	&	-	&	-	&	6000	&	6700	\\
J0659+1414	&	0.3848	&	14.1	&	14	&	6.5	&	55	&	18.4	&	1150	&	4000	&	7800	\\
J0754+3231	&	1.4423	&	40	&	40	&	8	&	49	&	12.2	&	5800	&	6500	&	7100	\\
J0826+2637	&	0.5306	&	19.5	&	19	&	73	&	620	&	5.8	&	56700	&	16000	&	27000	\\
J0837+0610	&	1.2737	&	12.9	&	12	&	89	&	1040	&	23.9	&	55400	&	71900	&	107000	\\
J0922+0638	&	0.4306	&	27.3	&	27	&	52	&	128	&	10.5	&	5200	&	9600	&	14300	\\
J0928+3037	&	2.0919	&	22	&	22	&	-	&	-	&	50	&	-	&	12000	&	13300	\\
J0943+4109	&	2.2294	&	21	&	21	&	8.6	&	-	&	-	&	-	&	8000	&	8300	\\
J0946+0951	&	1.0977	&	15.3	&	14	&	4	&	-	&	40.8	&	700	&	16000	&	33600	\\
J0953+0755	&	0.253	&	2.96	&	3	&	400	&	2030	&	9.5	&	54000	&	120000	&	174000	\\
J1136+1551	&	1.1879	&	4.84	&	5	&	257	&	1280	&	31.7	&	48000	&	216500	&	408000	\\
J1239+2453	&	1.3824	&	9.25	&	9	&	110	&	260	&	51.1	&	7000	&	131000	&	215000	\\
J1503+2111	&	3.314	&	3.26	&	3	&	1.3	&	-	&	26	&	1100	&	6000	&	12000	\\
J1532+2745	&	1.1248	&	14.7	&	15	&	13	&	94	&	25.7	&	4100	&	6900	&	11300	\\
J1538+2345	&	3.4494	&	14.9	&	14	&	-	&	-	&	-	&	-	&	8000	&	9300	\\
J1543+0929	&	0.7484	&	35	&	33	&	78	&	400	&	46.2	&	6500	&	6000	&	11900	\\
J1635+2418	&	0.4905	&	24.3	&	24	&	9.1	&	50	&	17.4	&	1400	&	5000	&	5900	\\
J1741+2758	&	1.3607	&	29.1	&	29	&	3	&	30	&	7	&	5800	&	3600	&	4100	\\
J1752+2359	&	0.409	&	36.2	&	36	&	3.5	&	40	&	4	&	4100	&	4300	&	5000	\\
J1758+3030	&	0.9472	&	35.1	&	35	&	8.9	&	60	&	27	&	2100	&	4600	&	5100	\\
J1813+4013	&	0.931	&	41.6	&	-	&	8	&	-	&	-	&	-	&	-	&	-	\\
J1844+1454	&	0.3754	&	41.5	&	42	&	20	&	-	&	9.3	&	5500	&	8000	&	10400	\\
J1907+4002	&	1.2357	&	31	&	31	&	23	&	-	&	58.5	&	3300	&	8000	&	8400	\\
J1921+2153	&	1.3373	&	12.4	&	12	&	57	&	1900	&	30.9	&	82200	&	48000	&	60000	\\
J1932+1059	&	0.2265	&	3.18	&	3	&	303	&	950	&	7.4	&	29100	&	16000	&	33100	\\
J1939+2134	&	0.0015	&	71	&	71	&	240	&	>2200	&	0.038	&	-	&	-	&	-	\\
J2018+2839	&	0.5579	&	14.2	&	14	&	314	&	260	&	14.9	&	9700	&	16000	&	18600	\\
J2022+2854	&	0.3434	&	24.6	&	24	&	71	&	?	&	12	&	13900	&	4300	&	7200	\\
J2037+1942	&	2.0743	&	36.9	&	36	&	2	&	-	&	-	&	-	&	3600	&	4300	\\
J2046+1540	&	1.1382	&	39.8	&	39	&	11.5	&	-	&	9.6	&	9300	&	2700	&	3500	\\
J2113+2754	&	1.2028	&	25.1	&	25	&	18	&	130	&	13	&	12000	&	24000	&	27300	\\
J2139+2242	&	1.0835	&	44.2	&	44	&	-	&	30	&	91	&	350	&	4800	&	5600	\\
J2157+4017	&	1.5252	&	71.1	&	70	&	105	&	200	&	38.6	&	7900	&	6900	&	7600	\\
J2212+2933	&	1.0045	&	74.5	&	74	&	6.3	&	50	&	51.2	&	1000	&	3400	&	3800	\\
J2234+2114	&	1.3587	&	35.1	&	35	&	2.6	&	35	&	43	&	1100	&	4800	&	5800	\\
J2305+3100	&	1.5758	&	49.6	&	49	&	24	&	-	&	17.4	&	14900	&	6900	&	7700	\\
J2317+2149 	&	1.4446	&	20.9	&	21	&	15	&	100	&	20.2	&	7150	&	8000	&	9800	\\
J2324-0530	&	0.8687	&	15	&	15	&	-	&	-	&	-	&	-	&	3600	&	7200	\\
		\hline
	\end{tabular}
	\label{tab:tab2}
\end{table*}

If no 102 MHz flux-density estimate is available, column (6) presents the 400 MHz flux density recalculated to 111 MHz assuming the spectral index 1.5. Note that, if there is no flux density available at 400 MHz, this usually means that the pulsar was recently detected at PRAO (\citeauthor{Tyulbashev2016} (\citeyear{Tyulbashev2016}), \citeauthor{Tyulbashev2017a} (\citeyear{Tyulbashev2017a})). Estimates of the expected, mean, and peak flux densities were obtained using the relation $S_{peak1}=S_{111}\times (P/W_{1/2})$. The periods, dispersion measures, flux densities, and pulse half-widths at 400 MHz were taken from the ATNF catalog. Note that the pulse half-width $W_{1/2}$ at 111 MHz can be $20\%$ larger than at 400 MHz (\citeauthor{Malov2010}, \citeyear{Malov2010}). It is of interest to compare the expected 111-MHz peak flux density $S_{peak1}$ and the corrected value $S_{peak3}$ recalculated from $S_{peak2}$. For pulses with durations shorter than the readout time, $S_{peak3}$ could be underestimated. The values $S_{peak2}$ were estimated based on the visible noise corridor and the assumption that the narrowest noise corridor corresponds to 1/7 of the threshold signal presented in Table~\ref{tab:tab1}. Therefore, the flux-density estimates $S_{peak2}$ and $S_{peak3}$ are fairly crude. The uncertainties in these estimates could reach $100\%$; we hope that they will be reduced to $20-30\%$ after taking into account variations in the gain coefficients between the frequency channels and the application of a calibration for all 128 BSA beams based on calibration sources. Systems for absolute (using calibration sources) and relative (using a noise signal with a known temperature) calibration of the flux densities are currently in the testing phase.

\begin{figure*}
\begin{center}
	\includegraphics[width=0.9\textwidth]{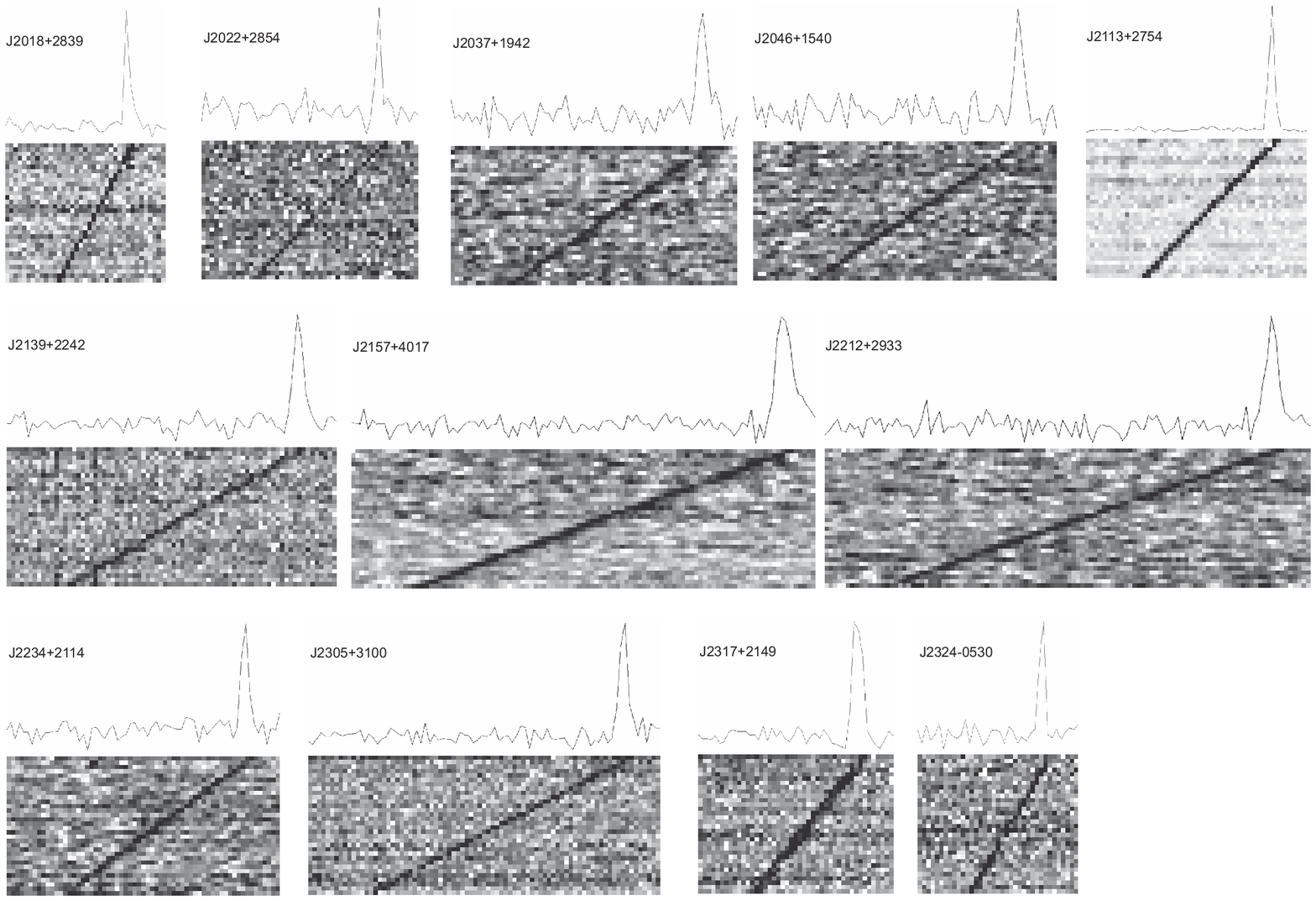}
    \caption{Contd.)}
    \label{fig:fig1btransients}
\end{center}
\end{figure*}

The minimum values of the measured peak flux density $S_{peak2}$ are 2.7 Jy, close to the expected threshold sensitivity for a search for individual flares using the BSA in Table~\ref{tab:tab1}. Reprocessing the observations with inter-channel gain calibration will increase the sensitivity by about $20-30\%$, especially at declinations $\delta<5^o$.

Let us note certain unexpected results associated with the detection of pulses from known pulsars. The pulsars J0528+2200 and J0534+2200 (the Crab pulsar) have close enough coordinates in both right ascension and declination that they could simultaneously fall into an BSA beam. Their dispersion measures are also similar, 51 and 57 pc/cm$^3$. The processed recordings correspond to the same areas on the sky, and their pulses are accordingly mixed. Therefore, it is difficult to distinguish these pulsars visually based on the appearance of the dynamical spectrum and the difference of their right ascensions. However, the pulse profile of the Crab pulsar shows obvious scattering, while no scattering is observed for J0528+2200. Note also that the period of the Crab pulsar is 33 ms, comparable to the time constant we used.

The pulsar J1939+2134 is one of the shortest period pulsars currently known ($P = 0.00156 s$). Its period is a factor of eight shorter than our readout interval. This would seem to suggest that we should not have detected this pulsar, but to our surprise it was detected in one set of observations. This could indicate that we have detected a new transient, whose coordinates and dispersion measure coincide with J1939+2134; however, it is more logical to suppose that we detected a rare giant pulse from this pulsar, especially given that this pulsar is known to give rise to giant pulses (\citeauthor{Wolszczan1984}, \citeyear{Wolszczan1984}). Taking into account the observations that are missing due to interference, we estimate that we saw one giant pulse among 2 500 000 pulses emitted by this pulsar over the four weeks of our monitoring.

\begin{figure*}
\begin{center}
	\includegraphics[width=1.0\textwidth]{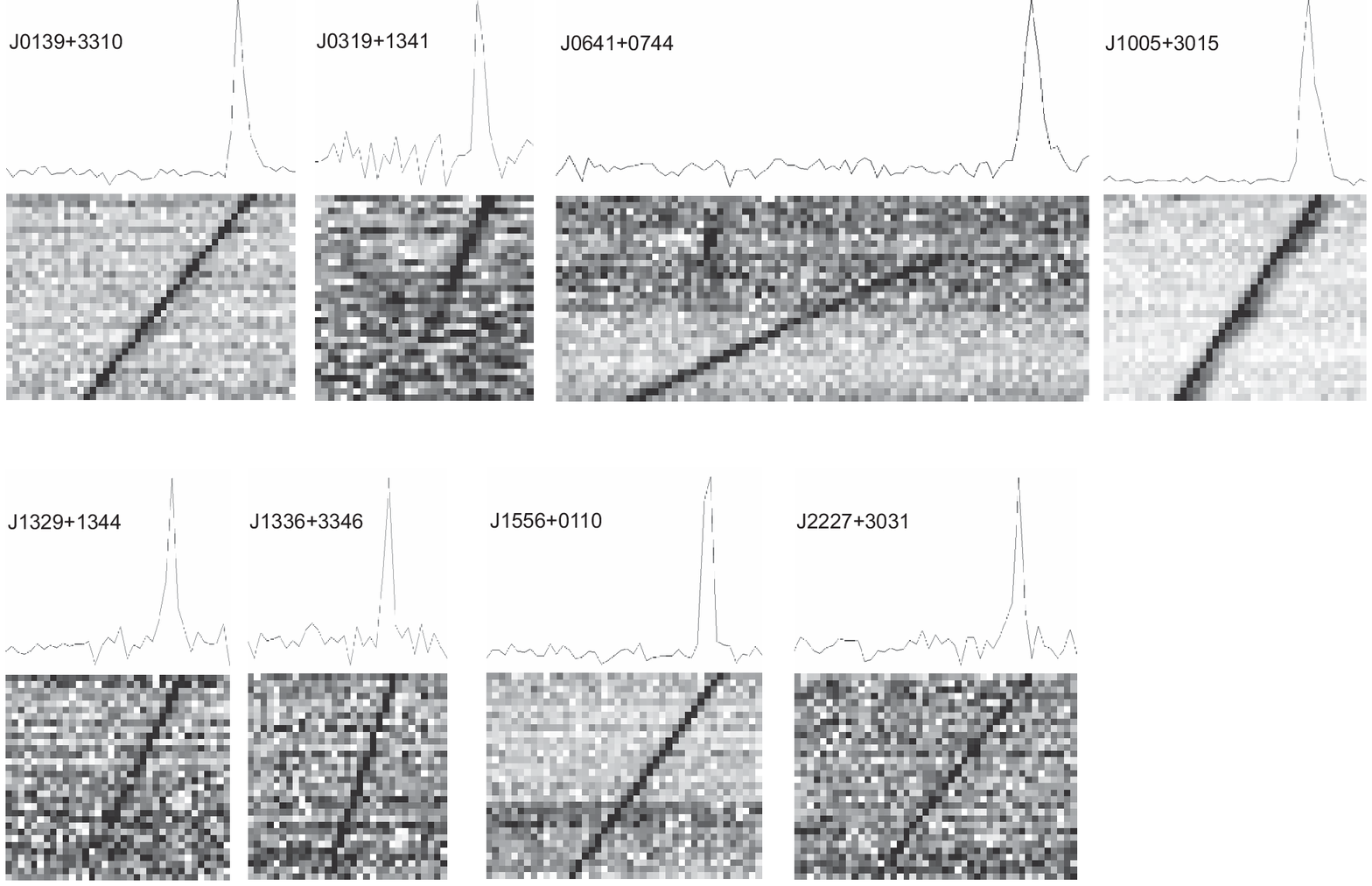}
    \caption{Profiles and dynamical spectra of individual pulses for detected fast transients.}
    \label{fig:fig2transients}
\end{center}
\end{figure*}

In addition to pulses from known pulsars, we also detected flares that could not be identified with pulsars in the ATNF catalog or objects in the RRATalog. We were able to detect pulses in eight directions on the sky. Fig.~\ref{fig:fig2transients} presents these pulses with the best SNR, together with their dynamical spectra.

\begin{table*}
	\centering
	\caption{Main characteristics of detected transients}
	\label{tab:tab3}
	\begin{tabular}{cccccc} 
		\hline
Name of transient & $\alpha_{2000} (^{h,m,s})$ & $\delta_{2000} (^{o,'})$ & $DM$ pc/cm$^3$ & $S_{peak4}$ (mJy) & N (number)\\
		\hline
J0139+3310	&	01 39 10	&	33 10 $\pm$ 15	&	$21\pm 2$	&	16000	&	14(+0)	\\
J0319+1341	&	03 19 30	&	13 41 $\pm$ 20	&	$12 \pm 2$	&	4500	&	2(+9)	\\
J0641+0744	&	06 41 10	&	07 44 $\pm$ 25	&	$52 \pm 4$	&	13000	&	1(+3)	\\
J1005+3015	&	10 05 30	&	30 15 $\pm$ 15	&	$17 \pm 2$	&	50000	&	4(+1)	\\
J1329+1344	&	13 29 00	&	13 44 $\pm$ 20	&	$12 \pm 2$	&	5800	&	1(+0)	\\
J1336+3346	&	13 36 00	&	33 46 $\pm$ 15	&	$8.5 \pm 1$	&	6000	&	1(+0)	\\
J1556+0110	&	15 56 00	&	01 08 $\pm$ 25	&	$19 \pm 3$	&	8800	&	1(+2)	\\
J2227+3031	&	22 27 00	&	30 31 $\pm$ 15	&	$19 \pm 2$	&	3900	&	13(+0)	\\
		\hline
	\end{tabular}
	\label{tab:tab3}
\end{table*}

Table~\ref{tab:tab3} presents information on the detected transients. Its columns contain (1) the J2000 name of the transient, (2)–(3) its right ascension (accuracy $\pm 2^m$) and declination (accuracy indicated in the table), (4) the estimated dispersion measure, (5) the estimated observed peak flux density of the pulse depicted in Fig.~\ref{fig:fig2transients}, and (6) the number of detections of the transient. The accuracy of the estimates in column (5) is low, and the peak flux density could be a factor of 2– 2.5 higher or lower than the value presented; this is due primarily to poor knowledge of the location of the transient relative to the BSA beams. Of the eight detected transients, only four were detected more than once. Therefore, we carried out a partial search using observations obtained in August 2015, reducing only areas with the coordinates of the transients in Table~\ref{tab:tab3}. The number of additional detections is shown in parentheses with a "+" sign.

Let us note certain properties of the new objects. Three pulses from the transient J2227+3031 were observed on September 3, and two pulses on September 2, 5, and 28. This enabled estimation of its period, $P = 0.8424$ s. A second inspection of the ATNF catalog led to the identification of PSR J2227+30 with $P = 0.842408$ s and $DM = 33$ pc/cm$^3$. We carefully verified our own estimated dispersion measure for the transient, which we found to be $19 \pm 2$ pc/cm$^3$. We suggest that the ATNF pulsar and the detected fast transient are the same object, and that the dispersion measure presented in Table~\ref{tab:tab3} is correct.

Inspection of recent publications shows that the strongest transients, J0139+3310 and J1005+3015, were apparently already detected at 111 MHz in earlier six-frequency BSA monitoring observations (\citeauthor{Samodurov2017} (\citeyear{Samodurov2017})). During the period from July 2012 through October 2013, 34 pulses at the coordinates $\alpha_{2000}= 1^h38^m30^s \pm 30^s$ and $\delta_{2000}=33^o41^\prime \pm 15^\prime$ and with $DM=21\pm 3$ pc/cm$^3$ were found; 12 pulses at the coordinates $\alpha_{2000}= 10^h05^m05^s \pm 30^s$ and $\delta_{2000}=30^o10^\prime \pm 15^\prime$ with $DM=18\pm 2$ pc/cm$^3$ were found. The remaining five transients have been detected for the first time.

\section{DISCUSSION}

In the previous section, we noted that the best threshold sensitivity in the transient search was 2.1 Jy, and that the typical value was appreciably worse. It is therefore not surprising that we were not able to detect known transients from the RRATalog. This catalog contains 45 objects whose declinations fall in the studied region of sky. Half of these are located at declinations $\delta < 5^o$, where the BSA
sensitivity is low. In addition, 10 of the 45 have $DM > 100$~pc/cm3$^3$, and the chance of detecting them at low frequencies is lower, due to the broadening of their pulses due to dispersion and scattering in the interstellar plasma. On the other hand, why were the transients we have detected missing from earlier surveys? Our estimates of $S_{peak4}$ for the pulses we have detected range from 3.9 Jy to 50 Jy. This corresponds to flux densities exceeding 0.7 Jy at 350 MHz (for a spectral index of 1.5), which is much higher than the threshold sensitivities in the transient surveys listed in Table~\ref{tab:tab1}. Therefore, the most probable situation is that the flare rate of the newly detected RRATs is low enough that the total observation time in the directions of these transients in the earlier surveys was not sufficient to enable detection of even one flare.

If this is the indeed the case, it demonstrates that the lower sensitivity of the BSA in transient searches is compensated by its ability to carry out longer monitoring. August 2017 marked three years of observations in the 32-channel regime. After rejecting observations due to interference, the accumulation time is equivalent to three days of observations at each point of the sky at the accessible declinations. These monitoring data should enable the detection of all RRAT objects whose flares are strong enough to exceed the BSA sensitivity threshold at least once in three days. If RRAT objects with low flare rates are encountered with the same probability as those with higher flare rates, the number of detected transients should be proportional to the observing time. In this case, reduction of three years of observations on the BSA could lead to the detection of 200–300 new flare sources with $SNR > 7$.

\section{CONCLUSION}

We have used data from a program of daily monitoring to search for strong ($SNR \ge 7$) pulses with DM = 3–100 pc/cm$^3$ within a region covering 15 000 deg$^2$ of the northern sky. Reduction of 28 days of observations in September 2015 led to the detection of 47 known pulsars and 7 transients. Two of these seven transients were observed earlier. The dispersion measures of the five new transients range from 9 to 55 pc/cm3$^3$, suggesting that these are Galactic objects. Five of the seven fast transients display repeated pulses, and are very like RRATs. \newline

{ACKNOWLEDGMENTS}

This work was supported by the Russian Foundation for Basic Research (grants 16-02-00954, 15-07-02830, 16-29-13074) and the program of the Presidium of the Russian Academy of Sciences “Transitional and Explosive Processes in Astrophysics”.

\bsp	
\label{lastpage}

\begin{thebibliography}{99}
\bibliographystyle{unsrt} 

\bibitem [McLaughlin(2006)]{McLaughlin2006} McLaughlin M.A., Lyne A.G., Lorimer D.~R., et al., 2006, Nature, 439, 817, doi:10.1038/nature04440

\bibitem [Shitov(2009)]{Shitov2009} Shitov Y.P., Kuzmin A.D., Dumskii D.V., Losovsky B.Y., 2009, ARep, 53, 561. doi:10.1134/S1063772909060080

\bibitem [Lorimer(2007)]{Lorimer2007} Lorimer D.R., Bailes M., McLaughlin M.A., et al., 2007, Science, 318, 777, doi:10.1126/science.1147532

\bibitem [Spitler(2016)]{Spitler2016} Spitler L.~G., Scholz P., Hessels J.~W.~T., Bogdanov S., Brazier A., Camilo F., Chatterjee S., et al., 2016, Natur, 531, 202. doi:10.1038/nature17168

\bibitem [Keane(2010)]{Keane2010} Keane E.F., 2010, PhD thesis, doi 10.1007/978-3-
642-19627-0

\bibitem [Tyulbashev(2017)]{Tyulbashev2017} Tyul'bashev S.~A., Tyul'bashev V.~S., 2017, ATsir, 1636

\bibitem [Shishov(2016)]{Shishov2016} Shishov V.~I., Chashei I.~V., Oreshko V.~V., Logvinenko S.~V., Tyul'bashev S.~A., Subaev I.~A., Svidskii P.~M., et al., 2016, ARep, 60, 1067. doi:10.1134/S1063772916110068

\bibitem [Tyulbashev(2016)]{Tyulbashev2016} Tyul'bashev S.A., Tyul'bashev V.S., Oreshko V.V., et al., 2016, Astronomy Reports, 60, 220, doi:10.1134/S1063772916020128

\bibitem [Tyulbashev(2017a)]{Tyulbashev2017a} Tyul'bashev S.~A., Tyul'bashev V.~S., Kitaeva M.~A., Chernyshova A.~I., Malofeev V.~M., Chashei I.~V., Shishov V.~I., et al., 2017, ARep, 61, 848. doi:10.1134/S1063772917100109

\bibitem [Malofeev(2000)]{Malofeev2000} Malofeev V.~M., Malov O.~I., Shchegoleva N.~V., 2000, ARep, 44, 436. doi:10.1134/1.163868

\bibitem [Karako-Argaman(2015)]{Karako-Argaman2015} Karako-Argaman C., Kaspi V.~M., Lynch R.~S., Hessels J.~W.~T., Kondratiev V.~I., McLaughlin M.~A., Ransom S.~M., et al., 2015, ApJ, 809, 67. doi:10.1088/0004-637X/809/1/67

\bibitem [Deneva(2016)]{Deneva2016} Deneva J.S., Stovall K., McLaughlin M. A.,
Bagchi M., et al., 2016, Astrophys. J. 821, 14

\bibitem [Kazantsev(2017)]{Kazantsev2017} Kazantsev A.~N., Potapov V.~A., 2017, ARep, 61, 747. doi:10.1134/S1063772917080054

\bibitem [Malov(2010)]{Malov2010} Malov O.~I., Malofeev V.~M., 2010, ARep, 54, 210. doi:10.1134/S1063772910030030

\bibitem [Wolszczan(1984)]{Wolszczan1984} Wolszczan A., Cordes J., Stinebring D., 1984, in Proceedings of an NRAOWorkshop, Green Bank,West Virginia, Ed. by S. P. Reynolds and D. R. Stinebring (Natl. Radio Astron. Observatory, 1984), p. 63

\bibitem [Samodurov(2017)]{Samodurov2017} Samodurov V. A., Pozanenko A. S., Rodin A. E., Churakov D. D., et al., 2017, Commun. Comput. Inform. Sci. Ser. 706, 130

\end{thebibliography}
\end{document}